\documentclass[twocolumn]{aastex63}

\usepackage{amsmath}

\newcommand{\lr}{\ifmmode{\lambda_{R_e}}\else{$\lambda_{R_e}$}\fi}
\newcommand{\lre}{\ifmmode{\lambda_{R_{\rm{e}}}}\else{$\lambda_{R_{\rm{e}}}$}\fi}
\newcommand{\kms}{\ifmmode{\,\rm{km}\, \rm{s}^{-1}}\else{$\,$km$\,$s$^{-1}$}\fi}
\newcommand{\update}[1]{#1}

\accepted{June 25, 2021}
\submitjournal{ApJ}

\shorttitle{Correlation Functions}
\shortauthors{Rutherford et al.}
\graphicspath{{./}{figures/}}

\begin{document}

\title{The SAMI Galaxy Survey: Detection of Environmental Dependence of Galaxy Spin in Observations and Simulations Using Marked Correlation Functions}

\email{trut2989@uni.sydney.edu.au}

\author{Tomas H. Rutherford}
\affiliation{Sydney Institute for Astronomy, School of Physics, A28, The University of Sydney, NSW, 2006, Australia}
\affiliation{ARC Centre of Excellence for All Sky Astrophysics in 3 Dimensions (ASTRO 3D), Australia}

\author{Scott M. Croom}
\affiliation{Sydney Institute for Astronomy, School of Physics, A28, The University of Sydney, NSW, 2006, Australia}
\affiliation{ARC Centre of Excellence for All Sky Astrophysics in 3 Dimensions (ASTRO 3D), Australia}

\author{Jesse van de Sande}
\affiliation{Sydney Institute for Astronomy, School of Physics, A28, The University of Sydney, NSW, 2006, Australia}
\affiliation{ARC Centre of Excellence for All Sky Astrophysics in 3 Dimensions (ASTRO 3D), Australia}

\author{Claudia del P. Lagos}
\affiliation{International Centre for Radio Astronomy Research (ICRAR), M468, University of Western Australia, 35 Stirling Hwy, Crawley, WA 6009, Australia}
\affiliation{ARC Centre of Excellence for All Sky Astrophysics in 3 Dimensions (ASTRO 3D), Australia}

\author{Joss Bland-Hawthorn}
\affiliation{Sydney Institute for Astronomy, School of Physics, A28, The University of Sydney, NSW, 2006, Australia}
\affiliation{ARC Centre of Excellence for All Sky Astrophysics in 3 Dimensions (ASTRO 3D), Australia}

\author{S. Brough}
\affiliation{School of Physics, University of New South Wales, NSW 2052, Australia}
\affiliation{ARC Centre of Excellence for All Sky Astrophysics in 3 Dimensions (ASTRO 3D), Australia}

\author{Julia J. Bryant}
\affiliation{Sydney Institute for Astronomy, School of Physics, A28, The University of Sydney, NSW, 2006, Australia}
\affiliation{Australian Astronomical Optics, AAO-USydney, School of Physics, University of Sydney, NSW 2006, Australia}
\affiliation{ARC Centre of Excellence for All Sky Astrophysics in 3 Dimensions (ASTRO 3D), Australia}

\author{Francesco D'Eugenio}
\affiliation{Sterrenkundig Observatorium, Universiteit Gent, Krijgslaan 281 S9, B-9000 Gent, Belgium}

\author{Matt S. Owers}
\affiliation{Department of Physics and Astronomy, Macquarie University, NSW 2109, Australia}
\affiliation{Astronomy, Astrophysics and Astrophotonics Research Centre, Macquarie University, Sydney, NSW 2109, Australia}



\begin{abstract}

The existence of a kinematic morphology-density relation remains uncertain, and instead stellar mass appears the more dominant driver of galaxy kinematics. We investigate the dependence of the stellar spin parameter proxy $\lambda_{R_e}$ on environment using a marked cross-correlation method with data from the SAMI Galaxy Survey. Our sample contains \update{710 galaxies} with spatially resolved stellar velocity and velocity dispersion measurements. By utilising the highly complete spectroscopic data from the GAMA survey, we calculate marked cross-correlation functions for SAMI galaxies using a pair count estimator and marks based on stellar mass and $\lambda_{R_e}$. We detect an anti-correlation of stellar kinematics with environment at the \update{3.2$\sigma$ level}, such that galaxies with low $\lambda_{R_e}$ values are preferably located in denser galaxy environments. However, a significant correlation between stellar mass and environment is also found (\update{correlation at 2.4$\sigma$}), as found in previous works. We compare these results to mock-observations from the cosmological EAGLE simulations, where we find a similar significant \lre\ anti-correlation with environment, and a mass and environment correlation. We demonstrate that the environmental correlation of \lre\ is not caused by the mass-environment relation. The significant relationship between \lre\ and environment remains when we exclude slow rotators. The signals in SAMI and EAGLE are strongest on small scales (10-100 kpc) as expected from galaxy interactions and mergers. Our work demonstrates that the technique of marked correlation functions is an effective tool for detecting the relationship between \lre\ and environment.
\end{abstract}



\section{Introduction} \label{sec:intro}

A relationship exists between galaxy properties and local environmental density \citep{DRESSLER}, such that early-type galaxies (ETGs) are preferably found in denser environments. Although the kinematic properties of galaxies do not correlate one-to-one with visual morphology, a tentative relationship between a proxy for the spin parameter, $\lambda_{R_e}$, and environmental density was presented by \cite{ATLAS3DVII}. This kinematic morphology-density relation (KMDR) suggests that the fraction  of slow rotating galaxies (low $\lambda_{R_e}$) increases towards denser environments. However, galaxy stellar mass also correlates with both environment and the intrinsic properties of galaxies. Thus, the question arises what the true physical driver of the KMDR is.

Further work supported the picture of a KMDR \citep{DEUGENIO2013, HOUGHTON2013,SCOTT2014,FOGARTY2014}. More recent results with larger galaxy samples find that the KMDR is driven mostly by stellar mass \citep{BROUGH2017,VEALE2017-2,GREENE2017}, but that the KMDR may also still exist at fixed stellar mass \citep{GRAHAM2019}. There is also evidence from simulations that points towards environmental dependence as a weaker secondary effect, but mass as the primary physical driver \citep[e.g.][]{LAGOS2017}. Some clear results have emerged, i.e. the trend with mass, but it is evident that any environmental dependence is likely to be a second order effect.

Multi-object integral field spectroscopy has revolutionised the number of galaxies with spatially resolved kinematic measurements. The Sydney-Australian-Astronomical-Observatory Multi-object Integral-Field Spectrograph (SAMI) galaxy survey has observed $\sim$3000 galaxies \citep{SAMII}, while the Mapping Nearby Galaxies at Apache Point Observatory (MaNGA) survey aims to observe $\sim$10,000 galaxies \citep{MANGAI}. Other ancillary surveys such as the Sloan Digital Sky Survey \citep[SDSS;][]{YORK2000} and the Galaxy and Mass Assembly Survey \citep[GAMA;][]{DRIVER2011} enable an accurate definition of environment, tracing the underlying large scale structure that exists in the Universe. 

With the growing wealth of spatially resolved kinematic data, the statistical tool of correlation functions becomes more powerful. It allows us to connect large scale structure in galaxy clustering to internal galaxy properties, in our case galaxy spin, parameterised by \lre\. Correlation functions have already demonstrated a relation between environment and the star-forming and morphological properties of galaxies \citep[e.g.][]{Madgwick,HERMIT1996}. Marked correlation functions, where galaxies are marked by some physical parameter, are even more effective at detecting and quantifying weak correlations with environment \citep{SHETH2004,HARKER2006},  making this method ideal for detecting a possible relation between $\lambda_{R_e}$ and environment.

\update{Most papers look at the fraction of fast and slow rotators as a function of mass and environment \citep{ATLAS3DVII,BROUGH2017,VEALE2017,GREENE2017,GRAHAM2019}, whereas a broader analysis of the \lre\ distribution as a function of mass and environment shows that environment might have a small impact on \lre\ \citep{WANG2020}.}

In this paper we aim to investigate the correlation between $\lambda_{R_e}$ and environmental density. We present an analysis using marked cross correlation functions applied to SAMI \citep{SAMII} and GAMA \citep{DRIVER2011} data, as well as mock observations from the EAGLE Simulations \citep{EAGLEI}. We adopt a $\Lambda$CDM cosmology, with $H_0=70\text{ km} \text{ s}^{-1} \text{ Mpc}^{-1}$, $\Omega_m=0.3$, $\Omega_{\Lambda}=0.7$.

\section{Observations and Simulations}
\label{sec:data}
\subsection{Observations}

The SAMI instrument \citep{SAMII} is mounted on the Anglo-Australian Telescope and provides a 1 degree diameter field of view. SAMI employs 13 fused fibre bundles \citep[Hexabundles;][]{BLANDHAWTHORN2011,BRYANT2014} with a high (75\%) fill factor. Each bundle contains 61 fibres of 1.6\arcsec diameter resulting in each IFU having a diameter of 15 \arcsec. The IFUs, as well as 26 sky fibres, are fed to the AAOmega spectrograph \citep{SHARP2006}, using the 580V grating at 3570-5750A giving a resolution of R=1808 ($\sigma$=70.4 \kms), and the R1000 grating from 6300-7400A giving a resolution of R-4304 ($\sigma$=29.6 \kms) \citep{VANDESANDE2017}. 

The SAMI Galaxy survey \citep{SAMII,BRYANT2015} selected galaxies from the GAMA \citep{DRIVER2011} survey, in addition to eight low-redshift clusters \citep{OWERS2017}. Reduced data cubes \citep{SHARP2015} and stellar kinematic maps are available with the SAMI Galaxy Survey data releases \citep{ALLEN2015,GREEN2018,SCOTT2018,CROOM2021}.

Our sample contains \update{1832 galaxies} with $\lambda_{R_e}$ measurements, derived from spatially resolved kinematic measurements as described in \cite{VANDESANDE2017}, and include an aperture correction \citep{VANDESANDE2017_MNRAS} and a seeing correction \citep{HARBORNE2020,VANDESANDE2020}. Furthermore, we define a volume-limited sample by selecting galaxies at: 1) $z < 0.06$ to avoid biases in the marked correlation function as the SAMI selection function results in different distributions of galaxy stellar masses in different volumes, and 2) $M/M_{\odot}>10^{10}$ to avoid low-completeness in the stellar kinematic sample. Alternatively, we could have treated each redshift section within the SAMI function selection individually, and taken a variance weighted mean of each resulting correlation function. While this achieves a stronger signal, the result becomes more difficult to interpret due to our ranking method for weights (Section \ref{sec:method}). The final sample contains \update{710 galaxies}, of all morphological types.\update{ This sample is unique amongst other similar surveys. For example, ATLAS$^{3\text{D}}$'s sample for kinematic analysis \citep{ATLAS3DVII} only contained 260 field ETGs, and MaNGA \citep[10,000 galaxies,][]{MANGAI} is only a narrow band in the mass redshift plane, with a complicated selection function that may make analyses that require volume limited samples more difficult to replicate.} 

Galaxies from the GAMA survey \citep{DRIVER2011,LISKE2015} serve as a background galaxy distribution for our analysis. GAMA adopted an r-band magnitude limit of $r_{\text{pet}} < 19.8$ mag. For our analysis, we restrict the GAMA sample at $z<0.06$ ( $z_{\text{max}}$ of SAMI) and define a volume-limited sample in redshift and $r$-band apparent magnitude. This was done so that the background distribution was not biased by a higher density of fainter galaxies at lower redshifts in an apparent magnitude limited sample. The GAMA data used in this paper came from three equatorial regions centred at $9^h$, $12^h$ and $14.5^h$ in RA, each of $12\times4$ deg$^2$. 

\subsection{Simulations}
\label{sec:sims}

We use galaxy mock-observations from the EAGLE hydrodynamical cosmological simulation suite \citep{EAGLEI} as presented by \cite{LAGOS2018}. A total of $7\times 10^8$ galaxies were extracted from the 100 Mpc$^3$ box, where each baryonic particle has an initial mass of $1.8\times 10^6$M$_{\odot}$, with a maximum gravitational softening length of 0.7 kpc. We adopt a stellar mass cut of M$_{\text{stars}} >5\times 10^9$ M$_{\odot}$, to ensure galaxies had angular momentum profiles that converged. $\lambda_{R_e}$ values were derived for mock-observations of this sample, leaving us with 5587 galaxies. A sample of 29737 galaxies not subjected to the stellar mass cut serves as the background galaxy distribution. These samples have an effective mass limit of $M/M_{\odot}>10^{8.5}$.

\begin{figure*}
 \includegraphics[width=\textwidth]{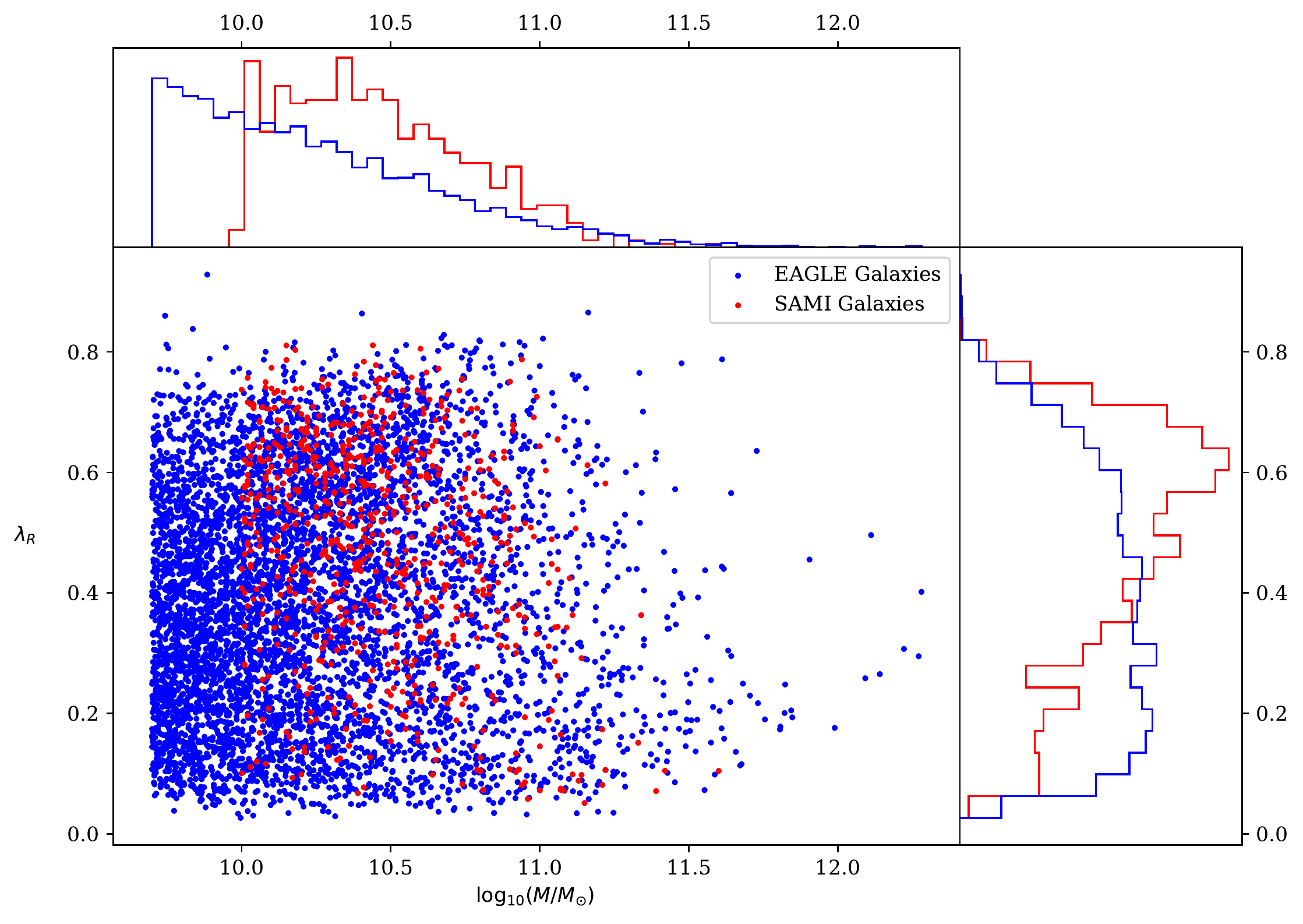}
 \caption{The distribution of SAMI (red) and EAGLE (blue) galaxies in $\text{log}_{10}(M/M_{\odot})$-$\lambda_{R_e}$ space. The normalised histograms of both samples for both parameters is also included. In this paper we adopt a mass limit for SAMI of $\log_{10}(M/M_{\odot})> 10$.}
 \label{fig:mass_spin_dist}
\end{figure*}

Additionally, as the mass distributions of SAMI and EAGLE galaxies are significantly different, we use a set of galaxies sampled from EAGLE in such a way to match the SAMI mass distribution. The initial distributions can be seen in Figure \ref{fig:mass_spin_dist}.

\section{Marked Correlation Functions}
\label{sec:method}
Marked statistics are a powerful tool to determine whether correlations between galaxy parameters depend on environment or not. Galaxies are assigned a mark, corresponding to some physical parameter, and a marked correlation function \citep{SHETH2004,SHETH2005} is calculated. We begin by defining the 2-point real-space correlation function $\xi(r)$, given by \cite{PEEBLES}:
\begin{equation}\label{eq:corr_func_def}
    dP=\rho^2[1+\xi(r)]dV_1dV_2,
\end{equation}
where $dP$ is the probability for two galaxies to be located at a distance $r$ from each other, in volume elements $dV_1$ and $dV_2$. $\rho$ is the mean density of galaxies in the volume considered. The correlation function $\xi(r)$ measures the ``overdensity''  of galaxies. In our case, we assume that redshift gives the radial distance to a galaxy. This is known as a redshift-space ($s$) correlation function, $\xi(s)$.

In practice, estimators are used to calculate $\xi(s)$. We use an estimator introduced by \cite{PEEBLES}:
\begin{equation}
    1+\xi(s)=\frac{\langle SG(s)\rangle}{\langle SG_R(s)\rangle},
\end{equation}
where $\langle SG(s)\rangle$ and $\langle SG_R(s)\rangle$ are pair counts between a SAMI ($S$) and GAMA ($G$) galaxy, and a SAMI and random GAMA ($G_R$) galaxy respectively. These pair counts are calculated by first defining radial bins, equally spaced in $\log_{10}(s)$. A SAMI galaxy is selected, and the distance to each GAMA galaxy is calculated. A count is then added to each relevant radial bin. This is repeated for all SAMI galaxies. Random GAMA galaxies are a sample of GAMA-like galaxies, created in such a way to match GAMA's selection function.

We now assign SAMI galaxies marks ($m$). Galaxies are ranked in $\lambda_{R_e}$ and log stellar mass, with the mark taken as the rank, to ensure an equivalent dynamical range for both marks. Marked correlation functions can then be thought of as a ratio of galaxy marks to the mean mark, $\overline{m}$, as a function of galaxy separation \citep{SHETH2005}:
\begin{align}\label{eq:marked_def}
    M(s) & \equiv \frac{\sum m(\boldsymbol{x})m(\boldsymbol{y})\mathcal{I}(|\boldsymbol{x}-\boldsymbol{y}|-s)}{\overline{m_x}\overline{m_y}\sum\mathcal{I}(|\boldsymbol{x}-\boldsymbol{y}|-s)}\nonumber\\
    &=\frac{\sum m(\boldsymbol{x})\mathcal{I}(|\boldsymbol{x}-\boldsymbol{y}|-s)}{\overline{m}\rho^2[1+\xi(s)]},
\end{align}
where $m(\boldsymbol{x})$ is the mark of a SAMI galaxy. We have defined the mark of all GAMA galaxies [$m(\boldsymbol{y})$] to be $1$, $\mathcal{I}(x)=0$ unless $x=0$, and the sum is over all SAMI-GAMA galaxy pairs. As we have divided by $\overline{m}$, $M(s)=1$ for all $s$ if no correlation between marks and environment exists.

We can also consider Equation \ref{eq:marked_def} in an alternate way. By a simple re-arrangement, the denominator can be expressed as one plus the regular correlation function, defined in Equation \ref{eq:corr_func_def}, and the numerator as one plus a ``weighted'' correlation function. This weighted correlation function, defined as $W(s)$, can be calculated using the same estimator as $\xi(s)$, except the $i^{\text{th}}$ SAMI galaxy contributes a weight of $m_i/\overline{m}$ to the relevant radial bins:
\begin{align}
    M(s)&\equiv \frac{\sum (m(\boldsymbol{x})/\overline{m})\mathcal{I}(|\boldsymbol{x}-\boldsymbol{y}|-s)}{\sum\mathcal{I}(|\boldsymbol{x}-\boldsymbol{y}|-s)}\equiv \frac{1+W(s)}{1+\xi(s)}\nonumber\\
    &=\frac{\langle WG(s) \rangle}{\langle SG(s) \rangle},
\end{align}
where $\langle WG(s) \rangle$ are \textit{weighted} pair counts between weighted SAMI ($W$) and GAMA ($G$) galaxies, and $\langle SG(s) \rangle$ are \textit{unweighted} pair counts between SAMI ($S$) and GAMA galaxies. Marked correlation functions with random marks are also calculated, as a check that any signal seen in the real functions is legitimate. Once our correlation functions are calculated, we use the Python emcee package \citep{EMCEE} to fit a function of the form:
\begin{equation}\label{eq:power_law}
    M(s)=1+As^{-m}
\end{equation}
We use this functional form as we expect $M(s)\approx 1$ for all scales other than small scales, where it may deviate according to possible spin-environment and mass-environment relations. Uncertainties are calculated from the $16^{\text{th}}$ and $84^{\text{th}}$ percentiles.
\subsection{Uncertainty Calculation}
\label{sec:errors} 
\update{We choose bootstrap re-sampling for our uncertainty estimate, as bootstrap re-sampling is robust and as shown by \cite{FISHER1994}, at worst overestimates error in correlation functions. Importantly, bootstrap uncertainties assume zero correlation between points, which is not strictly true in our case, as single galaxies contribute to multiple pair counts. However, on small scales the data points are largely independent, due to the small number of galaxies contributing to pairs at such small separation. As another test, we also evaluated another two different error estimations. We divided our sample into nine regions, and calculated a correlation function in each region, taking the standard deviation between the regions. We also calculated Poissonian errors for each bin. We found that the bootstrap errors were similar to Poissonian at small scales, and similar to the nine region standard deviation at large scales. Due to this, and our largely independent points at small scales, we use ordinary bootstrap re-sampling as a close approximation of errors.}

As bootstrap re-sampling generally overestimates uncertainties at small scales by a factor $\sqrt{3}$ \citep{CROOM1996}, the number of SAMI galaxies drawn per bootstrap sample was $3N$, where $N$ is the total number of galaxies, as also suggested by \cite{NORBERG2001}. Previous galactic correlation function works have used 10 re-samples \citep{HERMIT1996} or 20 re-samples \citep{Madgwick}. Due to our smaller data set in SAMI, we created 10,000 re-samples for each correlation function.

\section{Results}
\label{sec:results}
\subsection{SAMI Galaxies}
\begin{figure*}
 \includegraphics[width=\textwidth]{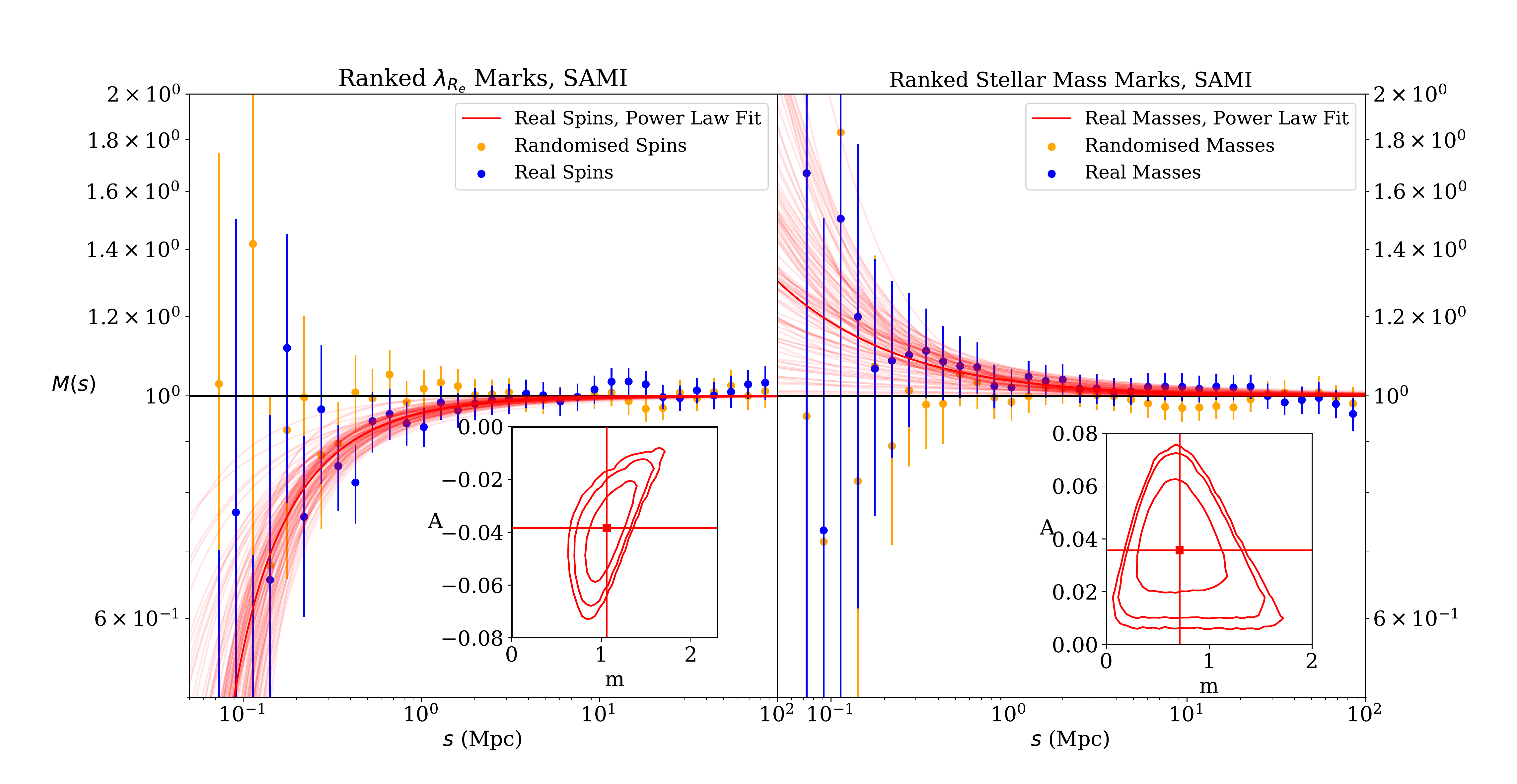}
 \caption{Marked correlation functions for ranked \lre\ (left panel) and ranked stellar mass marks (right panel), with SAMI galaxies (blue data). Randomised marks are plotted as a test of our method (orange data). We plot the best-fit power law (dark red line). One hundred samples from the MCMC chain (light red lines) are plotted, as a representation of the uncertainty in the fits. Contours for the 68$^{th}$, 90$^{th}$ and 95$^{th}$ percentiles are plotted for the real data in $A$ and $m$ space, as explored by the emcee algorithm, in the inset. The red overlay lines in these parameter spaces represent the values of $A$ and $m$ selected at the $50^{\text{th}}$ percentile. Measures of the significance can be found in Table \ref{tab:significance}. \lre\ is negatively correlated with environment, while stellar mass is positively correlated with environment in SAMI galaxies.}
  \label{fig:corr_funcs_sami}
\end{figure*}

We present marked correlation functions for SAMI Galaxies, with ranked $\lambda_{R_e}$ and stellar mass marks, in Figure \ref{fig:corr_funcs_sami}. 

Towards small scales $s$ we find that the marked correlation measurements (orange symbols) are significantly less than 1, i.e., there is a significant anti-correlation of ranked \lre\ with environment. The best-fit power-law (Equation \ref{eq:power_law}, Figure \ref{fig:corr_funcs_sami}a, red line), shows a significant turn below $M(s)=1$, beginning at around $s=1$ Mpc. \update{We find a best-fit value for $A=-0.038^{+0.012}_{-0.013}$, $3.2\sigma$ below zero}. This implies that galaxies with low $\lambda_{R_e}$ start being preferentially located in dense environments at scales of $s\approx 1$ Mpc. As done by \cite{HARKER2006}, we also take one large radial bin out to 1 Mpc to find significance at small scales ($M(<1Mpc)$). \update{We find $M(<1Mpc)=0.925\pm 0.034$, $2.2\sigma$ below 1}. Randomised marks are consistent with $M(s)=1$ at all scales.

In Figure \ref{fig:corr_funcs_sami}b we also detect a significant correlation of ranked stellar mass with environment. The best-fit power-law (red line) shows a significant up turn above $M(s)=1$ towards small scales in $s$, beginning at around $s=1$ Mpc. \update{We find a best-fit value of $A=0.036^{+0.015}_{-0.015}$, $2.4\sigma$ above zero}. This result implies that galaxies with high stellar mass start being preferentially located in dense environments at scales of $\approx 1$ Mpc. \update{We find $M(<1Mpc)=1.058\pm 0.042$, $1.4\sigma$ above zero}. This result is consistent with previous work, as galaxies are well known to cluster according to mass (or luminosity) \citep[e.g.][]{NORBERG2002}. Randomised marks are consistent with $M(s)=1$ at all scales.

\subsubsection{Further Analysis of SAMI Results}

\begin{figure*}
 \includegraphics[width=\textwidth]{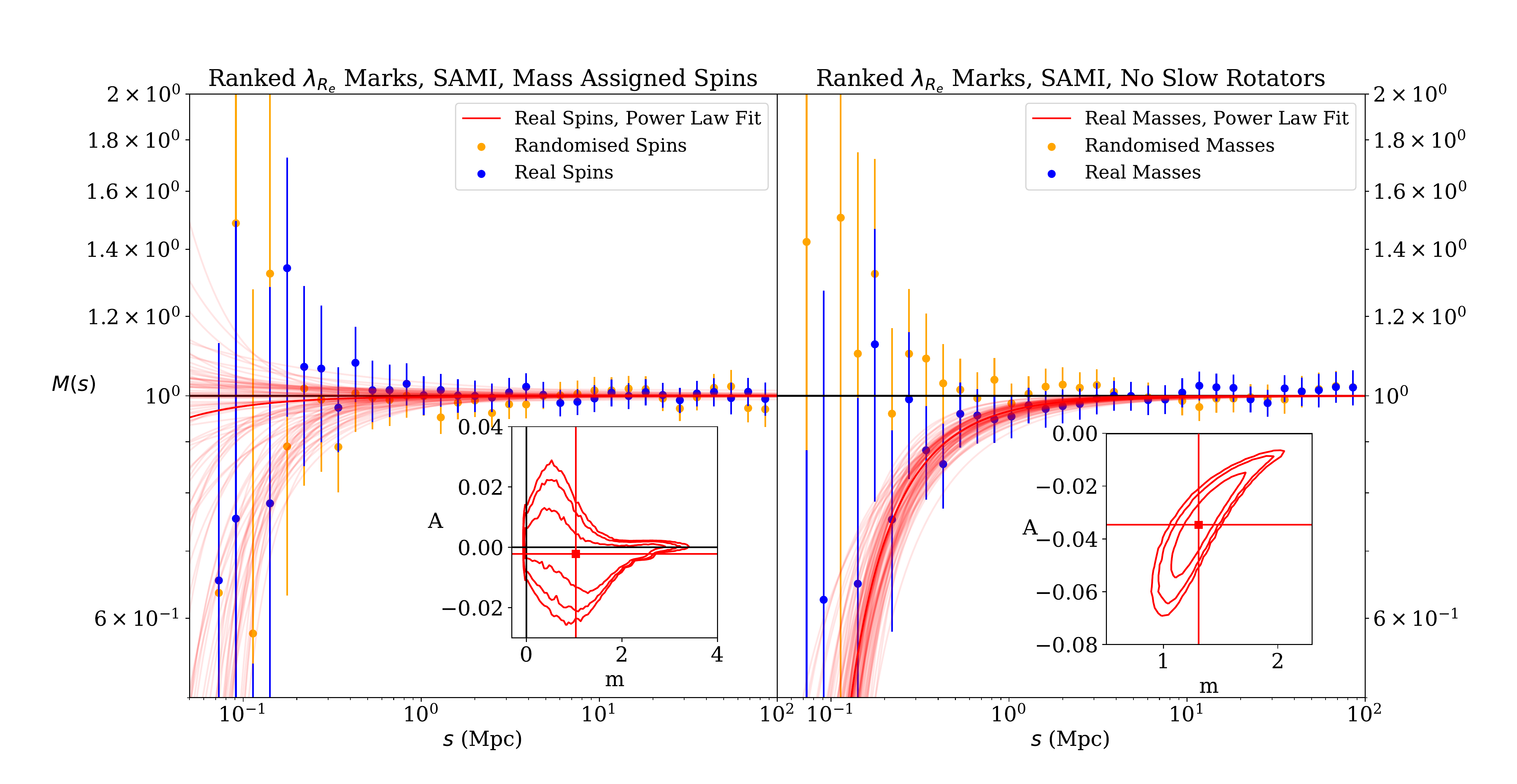}
 \caption{Marked correlation functions for different tests exploring whether our SAMI \lre\ anti-correlation with environment can be explained by mass trends. Selecting a galaxy's \lre\ value at random from 0.1 dex mass bins failed to reproduce a negative correlation (left panel). Removing slow rotators from the SAMI sample ($\lre<0.2$) still reproduces a negative correlation (right panel). Randomised marks are also plotted as a test of our method. Symbols and lines are similar as in Figure \ref{fig:corr_funcs_sami}, but now presented for mass-defined \lre\ and for a sample with slow rotators removed. The anti-correlation of \lre\ with environment is not driven purely by stellar mass, or by slow rotators.}
  \label{fig:randoms_and_noslowrot}
\end{figure*}

We have found an anti-correlation of \lre\ with environment, and a positive correlation of stellar mass with environment. We can compare these results with different tests. First, we reverse the rank order for \lre\ and compare $M(<1Mpc)$ for these reverse \lre\ ranks and stellar mass ranks. \update{We find $M(<1Mpc)=1.075\pm 0.043$ for reverse \lre\, and $M(<1Mpc)=1.058\pm0.041$ for stellar mass. This gives a difference of $0.017\pm0.059$}. Using this metric, mass and reverse \lre\ rank marks are consistent. However, this simplistic approach of assuming a one-to-one relationship between reversed \lre\ ranks and stellar mass ranks is not entirely supported, as there is a lot of scatter in this relationship \citep{CROOM2021}. In Figure \ref{fig:randoms_and_noslowrot}, we present two methods of testing for the physical driver of our SAMI $\lambda_{R_e}$-environment anti-correlation. Bins are defined in stellar mass of 0.1 dex in width. Galaxies are then assigned a \lre\ value from another random galaxy in their mass bin. This failed to reproduce a negative correlation with environment. The lack of correlation in this case implies that the signal we see between \lre\ and environment cannot be caused purely by a mass-environment relation. Removing slow rotators ($\lambda_{R_e}<0.2$) from the sample did reproduce a negative correlation with environment. This existence of a correlation implies that the observed correlation between \lre\ and environment cannot be attributed solely to slow rotators.

\subsection{EAGLE Galaxies}

We show the distribution in \lre\ and stellar mass of SAMI and EAGLE galaxies in Figure \ref{fig:mass_spin_dist}. As the mass distributions are significantly different, we create a set of observations sampled from EAGLE in such a way to match the SAMI mass distribution. Fifty bins spaced in $\log(M/M_{\odot})$ were defined across the mass range for SAMI, and EAGLE galaxies in each bin were drawn randomly until there were three times as many EAGLE galaxies in each bin as SAMI.

\begin{figure*}
 \includegraphics[width=\textwidth]{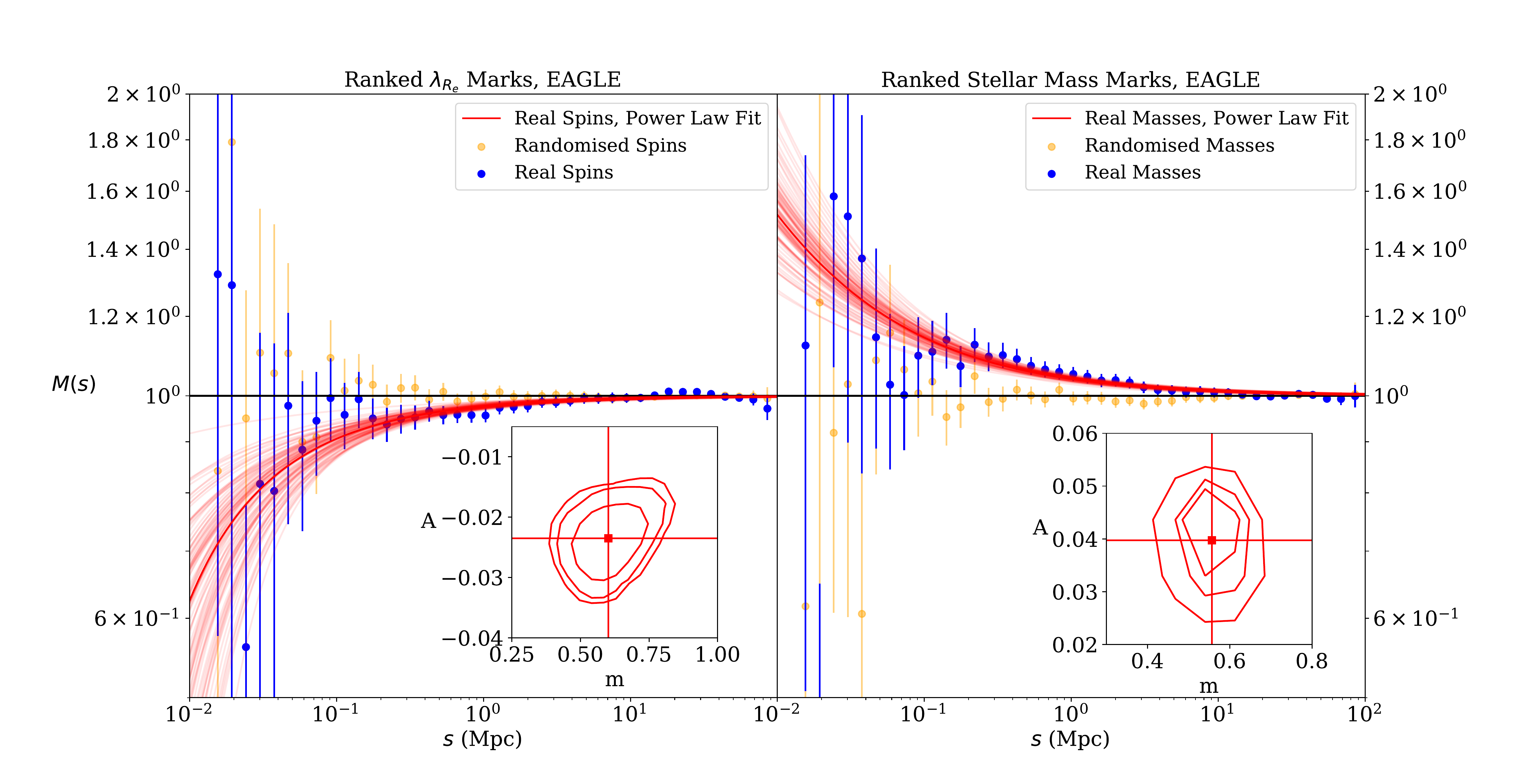}
 \caption{Marked correlation functions for ranked \lre\ (left panel) and ranked stellar mass marks (right panel). Symbols and lines are similar as in Figure \ref{fig:corr_funcs_sami}, but now presented for the EAGLE data. \lre\ is negatively correlated with environment, while stellar mass is positively correlated with environment in EAGLE galaxies.}
  \label{fig:corr_funcs_sim_resampled}
\end{figure*}

We present marked correlation functions for these re-sampled EAGLE galaxies, with ranked $\lambda_{R_e}$ and stellar mass marks, in Figure \ref{fig:corr_funcs_sim_resampled}. We see a significant anti-correlation of ranked $\lambda_{R_e}$ with environment in our marked correlation function for EAGLE (Figure \ref{fig:corr_funcs_sim_resampled}a). The best-fit power-law shows a significant downward trend towards small scales in $s$. $m$ is lower for EAGLE than for SAMI, so the correlation between \lre\ and environment extends to larger scales in EAGLE than in SAMI. We find a best-fit value of $A=-0.020^{+0.003}_{-0.003}$, $6.6\sigma$ below zero. This implies that in the EAGLE mock-observations, low-$\lambda_{R_e}$ galaxies are preferentially located in dense environments, out to scales of the order $\sim$1 Mpc. This is consistent with our SAMI results. We find $M(<1Mpc)=0.966\pm0.008$, $4.2\sigma$ below 1. Randomised marks are consistent with $M(s)=1$ at all scales.

We see a significant correlation of ranked stellar mass with environment in our marked correlation function for EAGLE. The best-fit power-law shows a significant upwards trend towards small scales in $s$. We find a best fit value of $A=0.030^{+0.003}_{-0.003}$, $10\sigma$ above zero. This implies that in the EAGLE mock-observations, high stellar mass galaxies are preferentially located in dense environments, out to scales of the order $\sim10$ Mpc. We find $M(<1Mpc)=1.066\pm0.015$, $4.4\sigma$ above 1. Randomised marks are consistent with $M(s)=1$ at all scales, as expected.

\begin{deluxetable*}{cclDl}\label{tab:significance}
\tablenum{1}
\tablecaption{Measurements of the significance of the signals in the correlation functions for SAMI and EAGLE data. We present the values of $A$ and $m$ for each data type and mark type, and the value of a single bin below 1 Mpc for each data type and mark type. The significance for the single bin is taken as the distance away from $M(s)=1$, the expected value given no correlation.}
\tablewidth{0pt}
\tablehead{
\colhead{Data Type} & \colhead{Mark Type} &  \colhead{$A$} &
\multicolumn2c{$m$} & \colhead{Single Bin Value} 
}
\decimalcolnumbers
\startdata
SAMI Spin Data & Real & $-0.038^{+0.012}_{-0.013}$ & $1.060^{+0.196}_{-0.179}$ & $0.925\pm0.035$ \\[2pt] 
  & Random & $-0.008^{+0.005}_{-0.008}$ & $1.850^{+0.412}_{-0.317}$ & $0.981\pm0.038$ \\[2pt]
SAMI Mass Data & Real & $0.036^{+0.015}_{-0.015}$ & $0.713^{+0.320}_{-0.255}$ & $1.058\pm0.041$ \\[2pt]
   & Random & $-0.003^{+0.009}_{-0.011}$ & $0.795^{+0.706}_{-0.512}$ & $1.019\pm0.039$ \\[2pt]
  Re-sampled EAGLE Spin Data & Real & $-0.020^{+0.003}_{-0.003}$ & $0.529^{+0.092}_{-0.094}$ & $0.966\pm 0.008$ \\[2pt]
   & Random & $-0.001^{+0.002}_{-0.003}$ & $0.530^{+0.480}_{-0.332}$ & $1.000\pm 0.008$ \\[2pt]
  Re-sampled EAGLE Mass Data & Real & $0.030^{+0.003}_{-0.003}$ & $0.488^{+0.045}_{-0.046}$ & $1.066\pm 0.015$ \\[2pt]
   & Random & $-0.003^{+0.003}_{-0.003}$ & $0.379^{+0.414}_{-0.225}$ & $1.005\pm 0.015$ \\[2pt]
  Mass Assigned Spin Data, SAMI & Real & $-0.002^{+0.009}_{-0.008}$ & $1.035^{+0.807}_{-0.641}$ & $0.984\pm 0.038$\\[2pt]
   & Random & $-0.014^{+0.011}_{-0.013}$ & $0.792^{+0.536}_{-0.412}$ & $0.989\pm 0.038$\\[2pt]
  No Slow Rotators Spin Data, SAMI & Real & $-0.035^{+0.013}_{-0.014}$ & $1.308^{+0.218}_{-0.183}$ & $0.937\pm 0.037$\\[2pt]
   & Random & $0.005^{+0.014}_{-0.009}$ & $0.812^{+0.662}_{-0.460}$ & $1.008\pm 0.041$
\enddata
\end{deluxetable*}

\section{Conclusion}
We investigate the existence of a kinematic morphology-density relation (KMDR) using the technique of marked correlation functions. Ranking galaxies by their spin (parametrised by $\lambda_{R_e}$) and stellar mass allows us to investigate whether the relation is driven purely by stellar mass or if some residual spin correlation exists. This is the first time that marked correlation functions have been applied to galactic spin. We apply our analysis to both the SAMI observations and EAGLE simulation mock-observations, allowing us to compare SAMI to simulated results. The GAMA dataset is used as the ``background'' in the SAMI analysis for measuring large scale structure, due to GAMA's highly complete spectroscopic data. We estimate uncertainties in all our correlation functions through bootstraps with 10,000 re-samples, and estimate significance by fitting power laws of the form $M(s)=1+As^{-m}$ and looking at single bins below 1 Mpc.

We find the following results:
\begin{itemize}
    \item Mass is positively correlated with environment. Fitting our ranked mass mark correlation function by a power law gave parameters $A=0.036^{+0.015}_{-0.015}$, and $m=0.713^{+0.320}_{-0.255}$. These results are consistent with previous work \citep[e.g.][]{NORBERG2002}.
    \item Spin (parametrised by $\lambda_{R_e}$) is negatively correlated with environment. Fitting our ranked $\lambda_{R_e}$ mark correlation function by a power law gave parameters $A=-0.038^{+0.012}_{-0.013}$, and $m=1.060^{+0.196}_{-0.179}$. This is consistent with previous work that found evidence of a KMDR \citep{ATLAS3DVII,DEUGENIO2013,HOUGHTON2013,SCOTT2014,FOGARTY2014}, and the presence of an environmental effect on spin \citep{CHOI2018}.
    \item Spin anti-correlation with environment is not driven solely by stellar mass. Defining bins in stellar mass of 0.1 dex, galaxies are assigned random \lre\ values corresponding to their mass bin. This failed to reproduce a negative correlation with environment.
    \item The anti-correlation between spin and environment is not driven purely by slow rotators. When slow rotators are removed from the SAMI sample ($\lambda_{R_e}<0.2$), the anti-correlation is found to still be present.
    \item Using mock-observations from the EAGLE simulations we find qualitatively consistent results as compared to observed data. Mass is correlated with environment, and spin is anti-correlated with environment with EAGLE galaxies as well as SAMI galaxies. A larger sample size in EAGLE allows us to see to larger scales than SAMI with high signal to noise, but the overall qualitative result remains the same. Although the relationship extends to larger scales for EAGLE, we know there are offsets between SAMI and EAGLE spin values \citep{VANDESANDE2019}. Investigation of whether these offsets can fully explain the larger scale result is beyond the scope of this paper.
\end{itemize}
    
Previous work has suggested that the $\lambda_{R_e}$-environment anti-correlation is simply a manifestation of the mass-environment correlation, caused by dynamical friction \citep{BROUGH2017,HOUGHTON2013,CAPPELLARI2016,VEALE2017-2,GREENE2017}. Our results suggest that the \lre\-environment anti-correlation is not as simple as a re-statement of the mass-environment correlation. Physical process such as environmental quenching, mergers and interactions may drive galaxies towards lower \lre\ values in denser environments (see also van de Sande et al., in Prep). The connection between mean stellar population age and environment \citep[e.g.][]{SCOTT2017} combined with the relation between mean stellar age and the dynamical thickness or intrinsic shape of galaxies \citep{vandesande2018} indicates that the star-formation history of a galaxy, its environment, and dynamical properties are closely connected. 

Future IFS surveys such as HECTOR \citep{BRYANT2016} will improve the statistics that we find here, in particular at small separation.

\section*{ACKNOWLEDGEMENTS}
The SAMI Galaxy Survey is based on observations made at the Anglo-Australian Telescope. SAMI was developed jointly by the University of Sydney and the Australian Astronomical Observatory (AAO). The SAMI input catalogue is based on data taken from the Sloan Digital Sky Survey, the GAMA Survey and the VST ATLAS Survey. The SAMI Galaxy Survey is supported by the Australian Research Council (ARC) Centre of Excellence ASTRO 3D (CE170100013) and CAASTRO (CE110001020), and other participating institutions. The SAMI Galaxy Survey website is http://sami-survey.org/

JvdS, JJB, JBH, SB, MSO acknowledge support of the ARC (DE200100461, FT180100231, FL140100278, FT140101166, FT140100255). The SAMI instrument was funded by the AAO and JBH through FF0776384, LE130100198. FDE acknowledges funding through the H2020 ERC Consolidator Grant 683184. 


\bibliography{paperbib}{}

\begin{thebibliography}{}
\expandafter\ifx\csname natexlab\endcsname\relax\def\natexlab#1{#1}\fi
\providecommand{\url}[1]{\href{#1}{#1}}
\providecommand{\dodoi}[1]{doi:~\href{http://doi.org/#1}{\nolinkurl{#1}}}
\providecommand{\doeprint}[1]{\href{http://ascl.net/#1}{\nolinkurl{http://ascl.net/#1}}}
\providecommand{\doarXiv}[1]{\href{https://arxiv.org/abs/#1}{\nolinkurl{https://arxiv.org/abs/#1}}}

\bibitem[{{Allen} {et~al.}(2015){Allen}, {Croom}, {Konstantopoulos}, {Bryant},
  {Sharp}, {Cecil}, {Fogarty}, {Foster}, {Green}, {Ho}, {Owers}, {Schaefer},
  {Scott}, {Bauer}, {Baldry}, {Barnes}, {Bland-Hawthorn}, {Bloom}, {Brough},
  {Colless}, {Cortese}, {Couch}, {Drinkwater}, {Driver}, {Goodwin},
  {Gunawardhana}, {Hampton}, {Hopkins}, {Kewley}, {Lawrence}, {Leon-Saval},
  {Liske}, {L{\'o}pez-S{\'a}nchez}, {Lorente}, {McElroy}, {Medling}, {Mould},
  {Norberg}, {Parker}, {Power}, {Pracy}, {Richards}, {Robotham}, {Sweet},
  {Taylor}, {Thomas}, {Tonini}, \& {Walcher}}]{ALLEN2015}
{Allen}, J.~T., {Croom}, S.~M., {Konstantopoulos}, I.~S., {et~al.} 2015,
  \mnras, 446, 1567, \dodoi{10.1093/mnras/stu2057}

\bibitem[{{Bland-Hawthorn} {et~al.}(2011){Bland-Hawthorn}, {Bryant},
  {Robertson}, {Gillingham}, {O'Byrne}, {Cecil}, {Haynes}, {Croom}, {Ellis},
  {Maack}, {Skovgaard}, \& {Noordegraaf}}]{BLANDHAWTHORN2011}
{Bland-Hawthorn}, J., {Bryant}, J., {Robertson}, G., {et~al.} 2011, Optics
  Express, 19, 2649, \dodoi{10.1364/OE.19.002649}

\bibitem[{Brough {et~al.}(2017)Brough, {van de Sande}, Owers, {d'Eugenio},
  Sharp, Cortese, Scott, Croom, Bassett, Bekki, {Bland-Hawthorn}, Bryant,
  Davies, Drinkwater, Driver, Foster, Goldstein, {L{\'o}pez-S{\'a}nchez},
  Medling, Sweet, Taranu, Tonini, Yi, Goodwin, Lawrence, \&
  Richards}]{BROUGH2017}
Brough, S., {van de Sande}, J., Owers, M.~S., {et~al.} 2017, 844, 59,
  \dodoi{10.3847/1538-4357/aa7a11}

\bibitem[{{Bryant} {et~al.}(2014){Bryant}, {Bland-Hawthorn}, {Fogarty},
  {Lawrence}, \& {Croom}}]{BRYANT2014}
{Bryant}, J.~J., {Bland-Hawthorn}, J., {Fogarty}, L.~M.~R., {Lawrence}, J.~S.,
  \& {Croom}, S.~M. 2014, \mnras, 438, 869, \dodoi{10.1093/mnras/stt2254}

\bibitem[{Bryant {et~al.}(2015)Bryant, Owers, Robotham, Croom, Driver,
  Drinkwater, Lorente, Cortese, Scott, Colless, Schaefer, Taylor,
  Konstantopoulos, Allen, Baldry, Barnes, Bauer, {Bland-Hawthorn}, Bloom,
  Brooks, Brough, Cecil, Couch, Croton, Davies, Ellis, Fogarty, Foster,
  Glazebrook, Goodwin, Green, Gunawardhana, Hampton, Ho, Hopkins, Kewley,
  Lawrence, {Leon-Saval}, Leslie, McElroy, Lewis, Liske,
  {L{\'o}pez-S{\'a}nchez}, Mahajan, Medling, Metcalfe, Meyer, Mould,
  Obreschkow, O'Toole, Pracy, Richards, Shanks, Sharp, Sweet, Thomas, Tonini,
  \& Walcher}]{BRYANT2015}
Bryant, J.~J., Owers, M.~S., Robotham, A. S.~G., {et~al.} 2015, 447, 2857,
  \dodoi{10.1093/mnras/stu2635}

\bibitem[{{Bryant} {et~al.}(2016){Bryant}, {Bland-Hawthorn}, {Lawrence},
  {Croom}, {Brown}, {Venkatesan}, {Gillingham}, {Zhelem}, {Content},
  {Saunders}, {Staszak}, {van de Sande}, {Couch}, {Leon-Saval}, {Tims},
  {McDermid}, \& {Schaefer}}]{BRYANT2016}
{Bryant}, J.~J., {Bland-Hawthorn}, J., {Lawrence}, J., {et~al.} 2016, in
  Society of Photo-Optical Instrumentation Engineers (SPIE) Conference Series,
  Vol. 9908, Ground-based and Airborne Instrumentation for Astronomy VI, ed.
  C.~J. {Evans}, L.~{Simard}, \& H.~{Takami}, 99081F,
  \dodoi{10.1117/12.2230740}

\bibitem[{{Bundy} {et~al.}(2015){Bundy}, {Bershady}, {Law}, {Yan}, {Drory},
  {MacDonald}, {Wake}, {Cherinka}, {S{\'a}nchez-Gallego}, {Weijmans}, {Thomas},
  {Tremonti}, {Masters}, {Coccato}, {Diamond-Stanic}, {Arag{\'o}n-Salamanca},
  {Avila-Reese}, {Badenes}, {Falc{\'o}n-Barroso}, {Belfiore}, {Bizyaev},
  {Blanc}, {Bland-Hawthorn}, {Blanton}, {Brownstein}, {Byler}, {Cappellari},
  {Conroy}, {Dutton}, {Emsellem}, {Etherington}, {Frinchaboy}, {Fu}, {Gunn},
  {Harding}, {Johnston}, {Kauffmann}, {Kinemuchi}, {Klaene}, {Knapen},
  {Leauthaud}, {Li}, {Lin}, {Maiolino}, {Malanushenko}, {Malanushenko}, {Mao},
  {Maraston}, {McDermid}, {Merrifield}, {Nichol}, {Oravetz}, {Pan}, {Parejko},
  {Sanchez}, {Schlegel}, {Simmons}, {Steele}, {Steinmetz}, {Thanjavur},
  {Thompson}, {Tinker}, {van den Bosch}, {Westfall}, {Wilkinson}, {Wright},
  {Xiao}, \& {Zhang}}]{MANGAI}
{Bundy}, K., {Bershady}, M.~A., {Law}, D.~R., {et~al.} 2015, \apj, 798, 7,
  \dodoi{10.1088/0004-637X/798/1/7}

\bibitem[{{Cappellari}(2016)}]{CAPPELLARI2016}
{Cappellari}, M. 2016, \araa, 54, 597,
  \dodoi{10.1146/annurev-astro-082214-122432}

\bibitem[{Cappellari {et~al.}(2011)Cappellari, Emsellem, Krajnovi{\'c},
  McDermid, Serra, Alatalo, Blitz, Bois, Bournaud, Bureau, Davies, Davis, {de
  Zeeuw}, Khochfar, Kuntschner, Lablanche, Morganti, Naab, Oosterloo, Sarzi,
  Scott, Weijmans, \& Young}]{ATLAS3DVII}
Cappellari, M., Emsellem, E., Krajnovi{\'c}, D., {et~al.} 2011, 416, 1680,
  \dodoi{10.1111/j.1365-2966.2011.18600.x}

\bibitem[{{Choi} {et~al.}(2018){Choi}, {Yi}, {Dubois}, {Kimm}, {Devriendt}, \&
  {Pichon}}]{CHOI2018}
{Choi}, H., {Yi}, S.~K., {Dubois}, Y., {et~al.} 2018, \apj, 856, 114,
  \dodoi{10.3847/1538-4357/aab08f}

\bibitem[{{Croom} \& {Shanks}(1996)}]{CROOM1996}
{Croom}, S.~M., \& {Shanks}, T. 1996, \mnras, 281, 893,
  \dodoi{10.1093/mnras/281.3.893}

\bibitem[{Croom {et~al.}(2012)Croom, Lawrence, {Bland-Hawthorn}, Bryant,
  Fogarty, Richards, Goodwin, Farrell, Miziarski, Heald, Jones, Lee, Colless,
  Brough, Hopkins, Bauer, Birchall, Ellis, Horton, {Leon-Saval}, Lewis,
  {L{\'o}pez-S{\'a}nchez}, Min, Trinh, \& Trowland}]{SAMII}
Croom, S.~M., Lawrence, J.~S., {Bland-Hawthorn}, J., {et~al.} 2012, 421, 872,
  \dodoi{10.1111/j.1365-2966.2011.20365.x}

\bibitem[{{Croom} {et~al.}(2021){Croom}, {Owers}, {Scott}, {Poetrodjojo},
  {Groves}, {van de Sande}, {Barone}, {Cortese}, {D'Eugenio}, {Bland-Hawthorn},
  {Bryant}, {Oh}, {Brough}, {Agostino}, {Casura}, {Catinella}, {Colless},
  {Cecil}, {Davies}, {Drinkwater}, {Driver}, {Ferreras}, {Foster},
  {Fraser-McKelvie}, {Lawrence}, {Leslie}, {Liske}, {L{\'o}pez-S{\'a}nchez},
  {Lorente}, {McElroy}, {Medling}, {Obreschkow}, {Richards}, {Sharp}, {Sweet},
  {Taranu}, {Taylor}, {Tescari}, {Thomas}, {Tocknell}, \&
  {Vaughan}}]{CROOM2021}
{Croom}, S.~M., {Owers}, M.~S., {Scott}, N., {et~al.} 2021, \mnras,
  \dodoi{10.1093/mnras/stab229}

\bibitem[{{D'Eugenio} {et~al.}(2013){D'Eugenio}, {Houghton}, {Davies}, \&
  {Dalla Bont{\`a}}}]{DEUGENIO2013}
{D'Eugenio}, F., {Houghton}, R.~C.~W., {Davies}, R.~L., \& {Dalla Bont{\`a}},
  E. 2013, \mnras, 429, 1258, \dodoi{10.1093/mnras/sts406}

\bibitem[{Dressler(1980)}]{DRESSLER}
Dressler, A. 1980, 236, 351, \dodoi{10.1086/157753}

\bibitem[{Driver {et~al.}(2011)Driver, Hill, Kelvin, Robotham, Liske, Norberg,
  Baldry, Bamford, Hopkins, Loveday, Peacock, Andrae, {Bland -Hawthorn},
  Brough, Brown, Cameron, Ching, Colless, Conselice, Croom, Cross, {de
  Propris}, Dye, Drinkwater, Ellis, Graham, Grootes, Gunawardhana, Jones, {van
  Kampen}, Maraston, Nichol, Parkinson, Phillipps, Pimbblet, Popescu, Prescott,
  Roseboom, Sadler, Sansom, Sharp, Smith, Taylor, Thomas, Tuffs, Wijesinghe,
  Dunne, Frenk, Jarvis, Madore, Meyer, Seibert, {Staveley-Smith}, Sutherland,
  \& Warren}]{DRIVER2011}
Driver, S.~P., Hill, D.~T., Kelvin, L.~S., {et~al.} 2011, 413, 971,
  \dodoi{10.1111/j.1365-2966.2010.18188.x}

\bibitem[{{Fisher} {et~al.}(1994){Fisher}, {Davis}, {Strauss}, {Yahil}, \&
  {Huchra}}]{FISHER1994}
{Fisher}, K.~B., {Davis}, M., {Strauss}, M.~A., {Yahil}, A., \& {Huchra}, J.
  1994, \mnras, 266, 50, \dodoi{10.1093/mnras/266.1.50}

\bibitem[{{Fogarty} {et~al.}(2014){Fogarty}, {Scott}, {Owers}, {Brough},
  {Croom}, {Pracy}, {Houghton}, {Bland-Hawthorn}, {Colless}, {Davies}, {Jones},
  {Allen}, {Bryant}, {Goodwin}, {Green}, {Konstantopoulos}, {Lawrence},
  {Richards}, {Cortese}, \& {Sharp}}]{FOGARTY2014}
{Fogarty}, L.~M.~R., {Scott}, N., {Owers}, M.~S., {et~al.} 2014, \mnras, 443,
  485, \dodoi{10.1093/mnras/stu1165}

\bibitem[{{Foreman-Mackey} {et~al.}(2013){Foreman-Mackey}, {Hogg}, {Lang}, \&
  {Goodman}}]{EMCEE}
{Foreman-Mackey}, D., {Hogg}, D.~W., {Lang}, D., \& {Goodman}, J. 2013, \pasp,
  125, 306, \dodoi{10.1086/670067}

\bibitem[{{Graham} {et~al.}(2019){Graham}, {Cappellari}, {Bershady}, \&
  {Drory}}]{GRAHAM2019}
{Graham}, M.~T., {Cappellari}, M., {Bershady}, M.~A., \& {Drory}, N. 2019,
  arXiv e-prints, arXiv:1910.05139.
\newblock \doarXiv{1910.05139}

\bibitem[{Green {et~al.}(2018)Green, Croom, Scott, Cortese, Medling, D'Eugenio,
  Bryant, {Bland-Hawthorn}, Allen, Sharp, Ho, Groves, Drinkwater, Mannering,
  {Harischand ra}, {van de Sande}, Thomas, O'Toole, McDermid, Vuong, Sealey,
  Bauer, Brough, Catinella, Cecil, Colless, Couch, Driver, Federrath, Foster,
  Goodwin, Hampton, Hopkins, Jones, Konstantopoulos, Lawrence, {Leon-Saval},
  Liske, {L{\'o}pez-S{\'a}nchez}, Lorente, Mould, Obreschkow, Owers, Richards,
  Robotham, Schaefer, Sweet, Taranu, Tescari, Tonini, \& Zafar}]{GREEN2018}
Green, A.~W., Croom, S.~M., Scott, N., {et~al.} 2018, 475, 716,
  \dodoi{10.1093/mnras/stx3135}

\bibitem[{{Greene} {et~al.}(2017){Greene}, {Leauthaud}, {Emsellem}, {Goddard},
  {Ge}, {Andrews}, {Brinkman}, {Brownstein}, {Greco}, {Law}, {Lin}, {Masters},
  {Merrifield}, {More}, {Okabe}, {Schneider}, {Thomas}, {Wake}, {Yan}, \&
  {Drory}}]{GREENE2017}
{Greene}, J.~E., {Leauthaud}, A., {Emsellem}, E., {et~al.} 2017, \apjl, 851,
  L33, \dodoi{10.3847/2041-8213/aa8ace}

\bibitem[{{Harborne} {et~al.}(2020){Harborne}, {van de Sande}, {Cortese},
  {Power}, {Robotham}, {Lagos}, \& {Croom}}]{HARBORNE2020}
{Harborne}, K.~E., {van de Sande}, J., {Cortese}, L., {et~al.} 2020, \mnras,
  497, 2018, \dodoi{10.1093/mnras/staa1847}

\bibitem[{{Harker} {et~al.}(2006){Harker}, {Cole}, {Helly}, {Frenk}, \&
  {Jenkins}}]{HARKER2006}
{Harker}, G., {Cole}, S., {Helly}, J., {Frenk}, C., \& {Jenkins}, A. 2006,
  \mnras, 367, 1039, \dodoi{10.1111/j.1365-2966.2006.10022.x}

\bibitem[{Hermit {et~al.}(1996)Hermit, Santiago, Lahav, Strauss, Davis,
  Dressler, \& Huchra}]{HERMIT1996}
Hermit, S., Santiago, B.~X., Lahav, O., {et~al.} 1996, 283, 709,
  \dodoi{10.1093/mnras/283.2.709}

\bibitem[{{Houghton} {et~al.}(2013){Houghton}, {Davies}, {D'Eugenio}, {Scott},
  {Thatte}, {Clarke}, {Tecza}, {Salter}, {Fogarty}, \&
  {Goodsall}}]{HOUGHTON2013}
{Houghton}, R.~C.~W., {Davies}, R.~L., {D'Eugenio}, F., {et~al.} 2013, \mnras,
  436, 19, \dodoi{10.1093/mnras/stt1399}

\bibitem[{Lagos {et~al.}(2018)Lagos, Schaye, Bah{\'e}, {Van de Sande}, Kay,
  Barnes, Davis, \& Dalla~Vecchia}]{LAGOS2018}
Lagos, C. d.~P., Schaye, J., Bah{\'e}, Y., {et~al.} 2018, 476, 4327,
  \dodoi{10.1093/mnras/sty489}

\bibitem[{{Lagos} {et~al.}(2017){Lagos}, {Theuns}, {Stevens}, {Cortese},
  {Padilla}, {Davis}, {Contreras}, \& {Croton}}]{LAGOS2017}
{Lagos}, C. d.~P., {Theuns}, T., {Stevens}, A. R.~H., {et~al.} 2017, \mnras,
  464, 3850, \dodoi{10.1093/mnras/stw2610}

\bibitem[{{Liske} {et~al.}(2015){Liske}, {Baldry}, {Driver}, {Tuffs},
  {Alpaslan}, {Andrae}, {Brough}, {Cluver}, {Grootes}, {Gunawardhana},
  {Kelvin}, {Loveday}, {Robotham}, {Taylor}, {Bamford}, {Bland-Hawthorn},
  {Brown}, {Drinkwater}, {Hopkins}, {Meyer}, {Norberg}, {Peacock}, {Agius},
  {Andrews}, {Bauer}, {Ching}, {Colless}, {Conselice}, {Croom}, {Davies}, {De
  Propris}, {Dunne}, {Eardley}, {Ellis}, {Foster}, {Frenk}, {H{\"a}u{\ss}ler},
  {Holwerda}, {Howlett}, {Ibarra}, {Jarvis}, {Jones}, {Kafle}, {Lacey},
  {Lange}, {Lara-L{\'o}pez}, {L{\'o}pez-S{\'a}nchez}, {Maddox}, {Madore},
  {McNaught-Roberts}, {Moffett}, {Nichol}, {Owers}, {Palamara}, {Penny},
  {Phillipps}, {Pimbblet}, {Popescu}, {Prescott}, {Proctor}, {Sadler},
  {Sansom}, {Seibert}, {Sharp}, {Sutherland}, {V{\'a}zquez-Mata}, {van Kampen},
  {Wilkins}, {Williams}, \& {Wright}}]{LISKE2015}
{Liske}, J., {Baldry}, I.~K., {Driver}, S.~P., {et~al.} 2015, \mnras, 452,
  2087, \dodoi{10.1093/mnras/stv1436}

\bibitem[{Madgwick {et~al.}(2003)Madgwick, Hawkins, Lahav, Maddox, Norberg,
  Peacock, Baldry, Baugh, {Bland-Hawthorn}, Bridges, Cannon, Cole, Colless,
  Collins, Couch, Dalton, De~Propris, Driver, Efstathiou, Ellis, Frenk,
  Glazebrook, Jackson, Lewis, Lumsden, Peterson, Sutherland, \&
  Taylor}]{Madgwick}
Madgwick, D.~S., Hawkins, E., Lahav, O., {et~al.} 2003, 344, 847,
  \dodoi{10.1046/j.1365-8711.2003.06861.x}

\bibitem[{Norberg {et~al.}(2001)Norberg, Baugh, Hawkins, Maddox, Peacock, Cole,
  Frenk, {Bland-Hawthorn}, Bridges, Cannon, Colless, Collins, Couch, Dalton,
  De~Propris, Driver, Efstathiou, Ellis, Glazebrook, Jackson, Lahav, Lewis,
  Lumsden, Madgwick, Peterson, Sutherland, \& Taylor}]{NORBERG2001}
Norberg, P., Baugh, C.~M., Hawkins, E., {et~al.} 2001, 328, 64,
  \dodoi{10.1046/j.1365-8711.2001.04839.x}

\bibitem[{Norberg {et~al.}(2002)Norberg, Baugh, Hawkins, Maddox, Madgwick,
  Lahav, Cole, Frenk, Baldry, {Bland -Hawthorn}, Bridges, Cannon, Colless,
  Collins, Couch, Dalton, De~Propris, Driver, Efstathiou, Ellis, Glazebrook,
  Jackson, Lewis, Lumsden, Peacock, Peterson, Sutherland, \&
  Taylor}]{NORBERG2002}
---. 2002, 332, 827, \dodoi{10.1046/j.1365-8711.2002.05348.x}

\bibitem[{Owers {et~al.}(2017)Owers, Allen, Baldry, Bryant, Cecil, Cortese,
  Croom, Driver, Fogarty, Green, Helmich, {de Jong}, Kuijken, Mahajan,
  McFarland, Pracy, Robotham, Sikkema, Sweet, Taylor, Verdoes~Kleijn, Bauer,
  {Bland -Hawthorn}, Brough, Colless, Couch, Davies, Drinkwater, Goodwin,
  Hopkins, Konstantopoulos, Foster, Lawrence, Lorente, Medling, Metcalfe,
  Richards, {van de Sande}, Scott, Shanks, Sharp, Thomas, \&
  Tonini}]{OWERS2017}
Owers, M.~S., Allen, J.~T., Baldry, I., {et~al.} 2017, 468, 1824,
  \dodoi{10.1093/mnras/stx562}

\bibitem[{Peebles(1980)}]{PEEBLES}
Peebles, P. J.~E. 1980, The Large-Scale Structure of the Universe

\bibitem[{{Schaye} {et~al.}(2015){Schaye}, {Crain}, {Bower}, {Furlong},
  {Schaller}, {Theuns}, {Dalla Vecchia}, {Frenk}, {McCarthy}, {Helly},
  {Jenkins}, {Rosas-Guevara}, {White}, {Baes}, {Booth}, {Camps}, {Navarro},
  {Qu}, {Rahmati}, {Sawala}, {Thomas}, \& {Trayford}}]{EAGLEI}
{Schaye}, J., {Crain}, R.~A., {Bower}, R.~G., {et~al.} 2015, \mnras, 446, 521,
  \dodoi{10.1093/mnras/stu2058}

\bibitem[{{Scott} {et~al.}(2014){Scott}, {Davies}, {Houghton}, {Cappellari},
  {Graham}, \& {Pimbblet}}]{SCOTT2014}
{Scott}, N., {Davies}, R.~L., {Houghton}, R. C.~W., {et~al.} 2014, \mnras, 441,
  274, \dodoi{10.1093/mnras/stu472}

\bibitem[{{Scott} {et~al.}(2017){Scott}, {Brough}, {Croom}, {Davies}, {van de
  Sande}, {Allen}, {Bland-Hawthorn}, {Bryant}, {Cortese}, {D'Eugenio},
  {Federrath}, {Ferreras}, {Goodwin}, {Groves}, {Konstantopoulos}, {Lawrence},
  {Medling}, {Moffett}, {Owers}, {Richards}, {Robotham}, {Tonini}, \&
  {Yi}}]{SCOTT2017}
{Scott}, N., {Brough}, S., {Croom}, S.~M., {et~al.} 2017, \mnras, 472, 2833,
  \dodoi{10.1093/mnras/stx2166}

\bibitem[{{Scott} {et~al.}(2018){Scott}, {van de Sande}, {Croom}, {Groves},
  {Owers}, {Poetrodjojo}, {D'Eugenio}, {Medling}, {Barat}, {Barone},
  {Bland-Hawthorn}, {Brough}, {Bryant}, {Cortese}, {Foster}, {Green}, {Oh},
  {Colless}, {Drinkwater}, {Driver}, {Goodwin}, {Gunawardhana}, {Federrath},
  {Harischandra}, {Jin}, {Lawrence}, {Lorente}, {Mannering}, {O'Toole},
  {Richards}, {Sanchez}, {Schaefer}, {Sealey}, {Sharp}, {Sweet}, {Taranu}, \&
  {Varidel}}]{SCOTT2018}
{Scott}, N., {van de Sande}, J., {Croom}, S.~M., {et~al.} 2018, \mnras, 481,
  2299, \dodoi{10.1093/mnras/sty2355}

\bibitem[{{Sharp} {et~al.}(2006){Sharp}, {Saunders}, {Smith}, {Churilov},
  {Correll}, {Dawson}, {Farrel}, {Frost}, {Haynes}, {Heald}, {Lankshear},
  {Mayfield}, {Waller}, \& {Whittard}}]{SHARP2006}
{Sharp}, R., {Saunders}, W., {Smith}, G., {et~al.} 2006, in Society of
  Photo-Optical Instrumentation Engineers (SPIE) Conference Series, Vol. 6269,
  Society of Photo-Optical Instrumentation Engineers (SPIE) Conference Series,
  ed. I.~S. {McLean} \& M.~{Iye}, 62690G, \dodoi{10.1117/12.671022}

\bibitem[{{Sharp} {et~al.}(2015){Sharp}, {Allen}, {Fogarty}, {Croom},
  {Cortese}, {Green}, {Nielsen}, {Richards}, {Scott}, {Taylor}, {Barnes},
  {Bauer}, {Birchall}, {Bland-Hawthorn}, {Bloom}, {Brough}, {Bryant}, {Cecil},
  {Colless}, {Couch}, {Drinkwater}, {Driver}, {Foster}, {Goodwin},
  {Gunawardhana}, {Ho}, {Hampton}, {Hopkins}, {Jones}, {Konstantopoulos},
  {Lawrence}, {Leslie}, {Lewis}, {Liske}, {L{\'o}pez-S{\'a}nchez}, {Lorente},
  {McElroy}, {Medling}, {Mahajan}, {Mould}, {Parker}, {Pracy}, {Obreschkow},
  {Owers}, {Schaefer}, {Sweet}, {Thomas}, {Tonini}, \& {Walcher}}]{SHARP2015}
{Sharp}, R., {Allen}, J.~T., {Fogarty}, L.~M.~R., {et~al.} 2015, \mnras, 446,
  1551, \dodoi{10.1093/mnras/stu2055}

\bibitem[{{Sheth} {et~al.}(2005){Sheth}, {Connolly}, \& {Skibba}}]{SHETH2005}
{Sheth}, R.~K., {Connolly}, A.~J., \& {Skibba}, R. 2005, arXiv e-prints, astro.
\newblock \doarXiv{astro-ph/0511773}

\bibitem[{{Sheth} \& {Tormen}(2004)}]{SHETH2004}
{Sheth}, R.~K., \& {Tormen}, G. 2004, \mnras, 350, 1385,
  \dodoi{10.1111/j.1365-2966.2004.07733.x}

\bibitem[{{van de Sande} {et~al.}(2017{\natexlab{a}}){van de Sande},
  {Bland-Hawthorn}, {Fogarty}, {Cortese}, {d'Eugenio}, {Croom}, {Scott},
  {Allen}, {Brough}, {Bryant}, {Cecil}, {Colless}, {Couch}, {Davies}, {Elahi},
  {Foster}, {Goldstein}, {Goodwin}, {Groves}, {Ho}, {Jeong}, {Jones},
  {Konstantopoulos}, {Lawrence}, {Leslie}, {L{\'o}pez-S{\'a}nchez}, {McDermid},
  {McElroy}, {Medling}, {Oh}, {Owers}, {Richards}, {Schaefer}, {Sharp},
  {Sweet}, {Taranu}, {Tonini}, {Walcher}, \& {Yi}}]{VANDESANDE2017}
{van de Sande}, J., {Bland-Hawthorn}, J., {Fogarty}, L. M.~R., {et~al.}
  2017{\natexlab{a}}, \apj, 835, 104, \dodoi{10.3847/1538-4357/835/1/104}

\bibitem[{{van de Sande} {et~al.}(2017{\natexlab{b}}){van de Sande},
  {Bland-Hawthorn}, {Brough}, {Croom}, {Cortese}, {Foster}, {Scott}, {Bryant},
  {d'Eugenio}, {Tonini}, {Goodwin}, {Konstantopoulos}, {Lawrence}, {Medling},
  {Owers}, {Richards}, {Schaefer}, \& {Yi}}]{VANDESANDE2017_MNRAS}
{van de Sande}, J., {Bland-Hawthorn}, J., {Brough}, S., {et~al.}
  2017{\natexlab{b}}, \mnras, 472, 1272, \dodoi{10.1093/mnras/stx1751}

\bibitem[{{van de Sande} {et~al.}(2018){van de Sande}, {Scott},
  {Bland-Hawthorn}, {Brough}, {Bryant}, {Colless}, {Cortese}, {Croom},
  {d'Eugenio}, {Foster}, {Goodwin}, {Konstantopoulos}, {Lawrence}, {McDermid},
  {Medling}, {Owers}, {Richards}, \& {Sharp}}]{vandesande2018}
{van de Sande}, J., {Scott}, N., {Bland-Hawthorn}, J., {et~al.} 2018, Nature
  Astronomy, 2, 483, \dodoi{10.1038/s41550-018-0436-x}

\bibitem[{{van de Sande} {et~al.}(2019){van de Sande}, {Lagos}, {Welker},
  {Bland-Hawthorn}, {Schulze}, {Remus}, {Bah{\'e}}, {Brough}, {Bryant},
  {Cortese}, {Croom}, {Devriendt}, {Dubois}, {Goodwin}, {Konstantopoulos},
  {Lawrence}, {Medling}, {Pichon}, {Richards}, {Sanchez}, {Scott}, \&
  {Sweet}}]{VANDESANDE2019}
{van de Sande}, J., {Lagos}, C. D.~P., {Welker}, C., {et~al.} 2019, \mnras,
  484, 869, \dodoi{10.1093/mnras/sty3506}

\bibitem[{{van de Sande} {et~al.}(2020){van de Sande}, {Vaughan}, {Cortese},
  {Scott}, {Bland-Hawthorn}, {Croom}, {Lagos}, {Brough}, {Bryant}, {Devriendt},
  {Dubois}, {D'Eugenio}, {Foster}, {Fraser-McKelvie}, {Harborne}, {Lawrence},
  {Oh}, {Owers}, {Poci}, {Remus}, {Richards}, {Schulze}, {Sweet}, {Varidel}, \&
  {Welker}}]{VANDESANDE2020}
{van de Sande}, J., {Vaughan}, S.~P., {Cortese}, L., {et~al.} 2020, arXiv
  e-prints, arXiv:2011.08199.
\newblock \doarXiv{2011.08199}

\bibitem[{{Veale} {et~al.}(2017){Veale}, {Ma}, {Greene}, {Thomas}, {Blakeslee},
  {McConnell}, {Walsh}, \& {Ito}}]{VEALE2017-2}
{Veale}, M., {Ma}, C.-P., {Greene}, J.~E., {et~al.} 2017, \mnras, 471, 1428,
  \dodoi{10.1093/mnras/stx1639}

\bibitem[{Veale {et~al.}(2017)Veale, Ma, Thomas, Greene, McConnell, Walsh, Ito,
  Blakeslee, \& Janish}]{VEALE2017}
Veale, M., Ma, C.-P., Thomas, J., {et~al.} 2017, 464, 356,
  \dodoi{10.1093/mnras/stw2330}

\bibitem[{{Wang} {et~al.}(2020){Wang}, {Cappellari}, {Peng}, \&
  {Graham}}]{WANG2020}
{Wang}, B., {Cappellari}, M., {Peng}, Y., \& {Graham}, M. 2020, \mnras, 495,
  1958, \dodoi{10.1093/mnras/staa1325}

\bibitem[{York {et~al.}(2000)York, Adelman, Anderson, Anderson, Annis, Bahcall,
  Bakken, Barkhouser, Bastian, Berman, Boroski, Bracker, Briegel, Briggs,
  Brinkmann, Brunner, Burles, Carey, Carr, Castander, Chen, Colestock,
  Connolly, Crocker, Csabai, Czarapata, Davis, Doi, Dombeck, Eisenstein,
  Ellman, Elms, Evans, Fan, Federwitz, Fiscelli, Friedman, Frieman, Fukugita,
  Gillespie, Gunn, Gurbani, {de Haas}, Haldeman, Harris, Hayes, Heckman,
  Hennessy, Hindsley, Holm, Holmgren, Huang, Hull, Husby, Ichikawa, Ichikawa,
  Ivezi{\'c}, Kent, Kim, Kinney, Klaene, Kleinman, Kleinman, Knapp, Korienek,
  Kron, Kunszt, Lamb, Lee, Leger, Limmongkol, Lindenmeyer, Long, Loomis,
  Loveday, Lucinio, Lupton, MacKinnon, Mannery, Mantsch, Margon, McGehee,
  McKay, Meiksin, Merelli, Monet, Munn, Narayanan, Nash, Neilsen, Neswold,
  Newberg, Nichol, Nicinski, Nonino, Okada, Okamura, Ostriker, Owen, Pauls,
  Peoples, Peterson, Petravick, Pier, Pope, Pordes, Prosapio, Rechenmacher,
  Quinn, Richards, Richmond, Rivetta, Rockosi, Ruthmansdorfer, {Sand ford},
  Schlegel, Schneider, Sekiguchi, Sergey, Shimasaku, Siegmund, Smee, Smith,
  Snedden, Stone, Stoughton, Strauss, Stubbs, SubbaRao, Szalay, Szapudi,
  Szokoly, Thakar, Tremonti, Tucker, Uomoto, Vanden~Berk, Vogeley, Waddell,
  Wang, Watanabe, Weinberg, Yanny, Yasuda, \& {SDSS Collaboration}}]{YORK2000}
York, D.~G., Adelman, J., Anderson, John~E., J., {et~al.} 2000, 120, 1579,
  \dodoi{10.1086/301513}

\end{thebibliography}
\bibliographystyle{aasjournal}



\end{document}